\documentclass[twocolumn,prb, showpacs]{revtex4}
\usepackage{graphics,graphicx}
\begin{document}
\title{Energetic Analysis of Magnetic Transitions in Ultra-small Nanoscopic Magnetic Rings}
\author{Deepak K.~Singh$^{1}$}
\author{Robert ~Krotkov$^{1}$}
\author{Mark T.~Tuominen$^{1*}$}
\affiliation{$^{1}$Department of Physics, University of
Massachusetts, Amherst, MA 01003}
\date{\today}

\begin{abstract}

In this article, we report on experimental and theoretical
investigations of magnetic transitions in  cobalt rings  of size
(diameter, width and thickness) comparable to the exchange length
of cobalt. Magnetization measurements were performed for two sets
of magnetic ring arrays: ultra-small magnetic rings (outer
diameter 13 nm, inner diameter 5nm and thickness 5 nm) and small
thin-walled magnetic rings (outer diameter 150 nm, width 5 nm and
thickness 5 nm). This is the first report on the fabrication and
magnetic properties of such small rings. Our calculations suggest
that if the magnetic ring's sizes are comparable to, or smaller
than, the exchange length of the magnetic material, then only two
magnetic states are important - the pure single domain state and
the flux closure vortex state. The onion-shape magnetic state does
not arise. Theoretical calculations are based on an energetic
analysis of pure and slightly distorted single domain and flux
closure vortex magnetic states. Based on the analytical
calculations, a phase diagram is also derived for ultra-small ring
structures exhibiting the region for vortex magnetic state
formations as a function of material parameter.

\end{abstract}

 \maketitle

\section{INTRODUCTION}
In recent times, the magnetic ring geometry has been extensively
studied, mostly because of its possible applications in magnetic
memory devices. The application in memory devices is mostly driven
by the fact that near zero field value, a narrow nanoscopic
magnetic ring forms a flux closure vortex magnetic
state.\cite{Yoo}$^{,}$ \cite{Klaui1}$^{,}$ \cite{Natali} A
magnetic ring in vortex state has zero total magnetization and
therefore each ring in an array acts like an individual memory
element. Ring geometry is used in atleast one current design of
magnetic random access memory (MRAM).\cite{Zhu1}

In addition to the vortex magnetic state, a small magnetic ring
also forms two other stable magnetic states: "onion state",
characterized by the presence of two head-to-head domain walls,
and single domain (SD) state, as shown in Fig. 1.\cite{Rothman}
However, we have found that if the ring sizes are sufficiently
small then onion magnetic state has higher energy and does not
arise. Therefore, magnetic transition processes in such
ultra-small rings involve only two magnetic states: saturating SD
state and flux closure vortex state. These conclusions are based
on the calculation of total magnetic energies for various possible
magnetic states and on the measurement of magnetization as a
function of applied field.

Because of the circular geometry of rings, shape anisotropy is
absent and also if the parent magnetic material is of
polycrystalline origin then magnetocrystalline anisotropy is
limited to random grains and can be ignored also.\cite{Skomsky}
Therefore in zero field the only competing energy terms in the
case of polycrystalline magnetic rings are magnetostatic energy
and quantum mechanical exchange energy.\cite{Cowburn1}$^{,}$
\cite{Aharoni}. Exchange energy favors the parallel alignment of
spins while the magnetostatic energy favors the circular
magnetization.

\begin{figure}
\centering
\includegraphics[width=8.0cm]{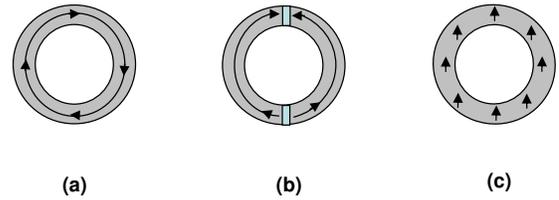} \vspace{-2mm}
\caption{(color online) Magnetic states of nano-rings. (a) Vortex
magnetic state. (b) Onion magnetic state, with domain walls. (c)
Single domain magnetic state.
} \vspace{-4mm}
\end{figure}

If the ring's sizes (diameter, thickness and width) are comparable
to the characteristic length (exchange length) of the parent
magnetic material then the magnetic transition processes are
expected to be different from those in relatively larger size
nanoscopic rings ($\sim$ 100 nm). In this new geometrical regime,
it is reasonable to assume that the magnetization vector remains
confined in the plane of the ring. In this article, we are
presenting the study of magnetization and magnetic transitions in
ultra-small polycrystalline magnetic, Co, rings of sizes
comparable to the exchange length of Co ($l$$_{ex}$ = 3.8
nm).\cite{Coey} For the first time, magnetic rings are studied at
such an unprecedented small length scale.

The magnetic rings are fabricated using copolymer template,
angular metal evaporation and ion-beam etching technique. Using
this fabrication technique, we have been able to fabricate arrays
of rings at two geometrical scales: ultra-small rings with outer
diameter 13 nm, ring width 4 nm and thickness $\sim$ 5 nm and
small rings with outer diameter 150 nm, ring width 5 nm and
thickness $\sim$ 5 nm. The latter are small rings with a thin
wall. Magnetization measurements are carried out for arrays of
both ultra-small and small rings. Experimental data are explained
by detailed theoretical calculations for these rings. The
theoretical calculations are based on the energetic analysis of
possible magnetic states with the underlying assumption that only
the lowest energy magnetic states will be excited. Different
magnetic states in a ring structure are obtained using reasonable
models of magnetization distortion on the ring's
circumferences.\cite{Deepak1}

\section{FABRICATION PROCEDURE}
Recently small ferromagnetic rings have been fabricated by
electron beam lithography,\cite{Yoo}$^{,}$ \cite{Heyderman}
evaporation over spheres\cite{Zhu2} and other
methods.\cite{Hobbs}$^{,}$ \cite{Kosiorek} Our nanoring
fabrication technique is described in detail in an earlier
work.\cite{Deepak2} The fabrication process for both ultra-small
and small rings with thin wall (arms) involve the creation of
nanoporous polymer templates, angular deposition of desired
material, Co, onto the wall of the pores and ion-beam etching
technique to remove undesired material from the top as well as
bottom of the template. In the case of ultra-small rings, the
template is created from a self-assembled diblock copolymer
film\cite{Albretch} while for small rings, the template is created
by electron-beam lithography technique\cite{Bal} out of the 40 nm
thick copolymer film PMMA [poly(methyl-metha-acrylate)]. The
diblock copolymer film has the pore size of 13 nm on the average
and thickness 36 nm. The lattice separation between pores is about
28 nm. In the case of copolymer PMMA film, the pore size is 150 nm
and the separation between pores (center-to-center distance) is
350 nm. The angular deposition of desired material onto the wall
of the nanopores is the most critical step of this fabrication
scheme and it depends on the critical deposition angle,
$\theta$$_{c}$. In our experiment $w$ = 13 nm and $h$ = 36 nm for
the fabrication of ultra-small rings and $w$ = 150 nm and $h$ = 40
nm for small rings, resulting in the critical angles of
$\theta$$_{c}$ $\sim$ 20$^{o}$ and 75$^{o}$ respectively. Angles
of $\theta$ = 23$^{o}$ and 75$^{o}$ were chosen for the
fabrication of ultra-small and small rings respectively. The
thickness of deposited Co material on nanopores varies in the
range of 4-5 nm (based on QCM reading). After material deposition,
calibrated ion-beam etching is used to get the desired thickness
of both ultra-small and small rings. The desired resulting
thicknesses of both ultra-small and small rings are $\sim$ 5 nm.

After ion-beam etching, the small ring samples are rinsed in
acetone solvent to remove the remaining polymer residues and are
characterized by field emission scanning electron microscope
(FESEM) (Figure 2). Structural characterization of ultra-small
rings is done using tunneling electron microscope (TEM) (Figure
2). Sample preparation of ultra-small rings for TEM imaging
involved the transfer of ring templates onto an electron
transparent substrate. It is a difficult process and can be found
in detail somewhere else\cite{Deepak2}. For the magnetic
measurement process, ultra-small rings were not transferred on
electron transparent substrate.

\begin{figure}
\centering
\includegraphics[width=8.0cm]{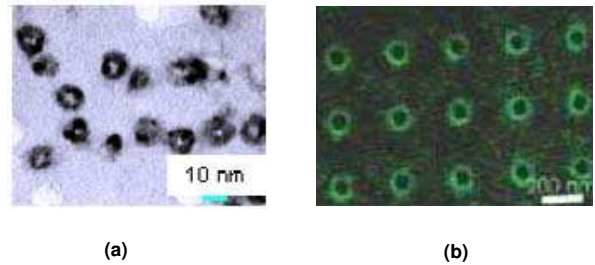} \vspace{-2mm}
\caption{(color online) Images of rings. (a) TEM image of
ultra-small rings and some empty diblock pores (rings came out of
these pores during sample preparation for imaging). (b) FESEM
images of small rings with thin wall.
} \vspace{-4mm}
\end{figure}

\section{MAGNETIZATION MEASUREMENTS AND DATA ANALYSIS}
Magnetic measurements of ultra-small and small ring arrays were
carried out in a SQUID magnetometer with base temperature 1.8 K
and field applied in-plane of the ring. Figure 3 and 4 shows
magnetization measurements for arrays of ultra-small and small
rings respectively, after the subtraction of a linear diamagnetic
background. As we can see in the low temperature in-plane
magnetization data for both ultra-small and small rings, the width
of the hysteresis curve is smaller near zero field value than near
the saturation which is similar to the magnetic transition in
nanoscopic narrow rings.\cite{Klaui2} In relatively larger size
nanoscopic rings, it is observed that the magnetic state is single
domain (SD) state at saturating field value and as the field is
decreased, the ring's magnetic state forms an "onion-shape", near
zero field value, and finally transforms to flux closure vortex
state (Figure 1).\cite{Natali}$^{,}$ \cite{Zhu1} It also depends
on the ring's width, diameter and ferromagnetic exchange
length.\cite{Li}

To have a complete understanding of these magnetic transition
phenomena, one needs to solve the following integro-differential
equation:\cite{Brown}$^{,}$ \cite{Aharoni2}
\begin{eqnarray*}
{l}_{ex}^{2}\frac{d^{2}{\beta}}{d^{2}{\phi}}+cos({\beta}){\rho}^{2}{h}_{M}-sin({\beta}){\rho}^{2}{h}_{M}-sin({\beta}){\rho}^{2}{h}_{ax}&=&0
\end{eqnarray*}
\begin{eqnarray}
{h}_{M}({\rho},{\phi})&=&{\int}{d}{\phi}^{'}{\sigma}({\beta},{\phi}^{'}){K}({\rho},{\phi},{\phi}^{'})
\end{eqnarray}
 where $\rho$ is the radial co-ordinate, $\sigma$ is magnetic
pole density (depending on  azimuthal position $\phi$$^{'}$), the
kernel $K$ is an explicit but complicated function involving
$\rho$ and $\phi$ and $X$ is the direction of applied magnetic
field. By solving these integro-differential equations, we can get
the functional $\beta$($h$, $\phi$) and since $m$($h$) =
$cos$$\beta$, explained later, (here we have taken $m$ = 1 just
for the sake of simplicity), so we can get $m$ (magnetization) as
a function of $h$ (dimensionless magnetic field). This is not an
easy task. Another approach is reverse approach.

\begin{figure}
\centering
\includegraphics[width=8.0cm]{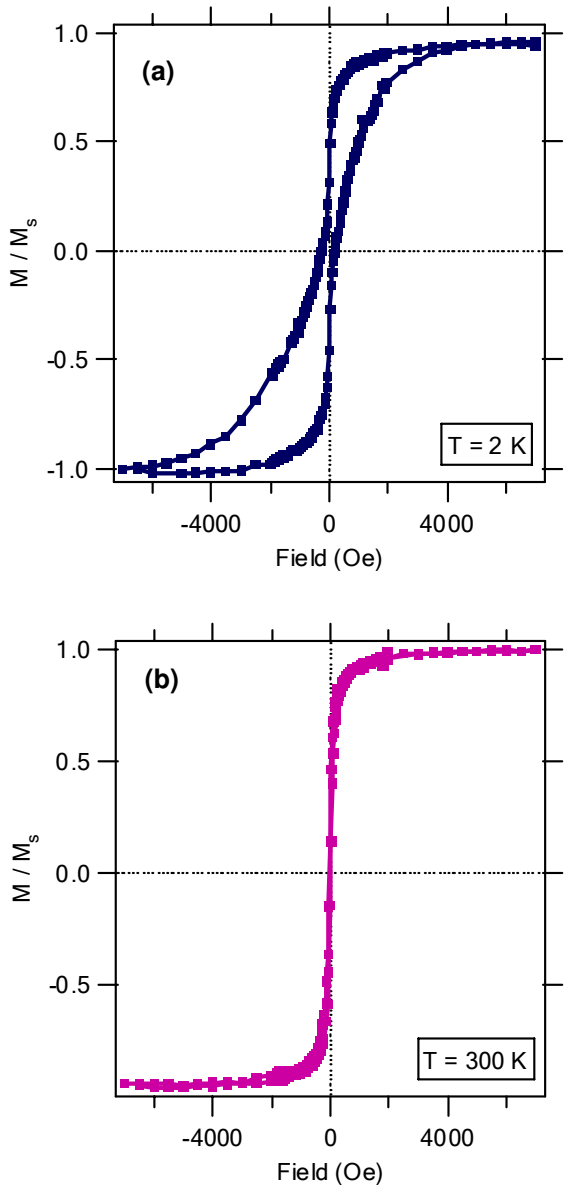} \vspace{-2mm}
\caption{(color online) In-plane magnetization measurements for an
array of ultrasmall Co rings. (a) Measurements at 2 K. (b)
Measurement at 300 K.
} \vspace{-4mm}
\end{figure}

\begin{figure}
\centering
\includegraphics[width=8.0cm]{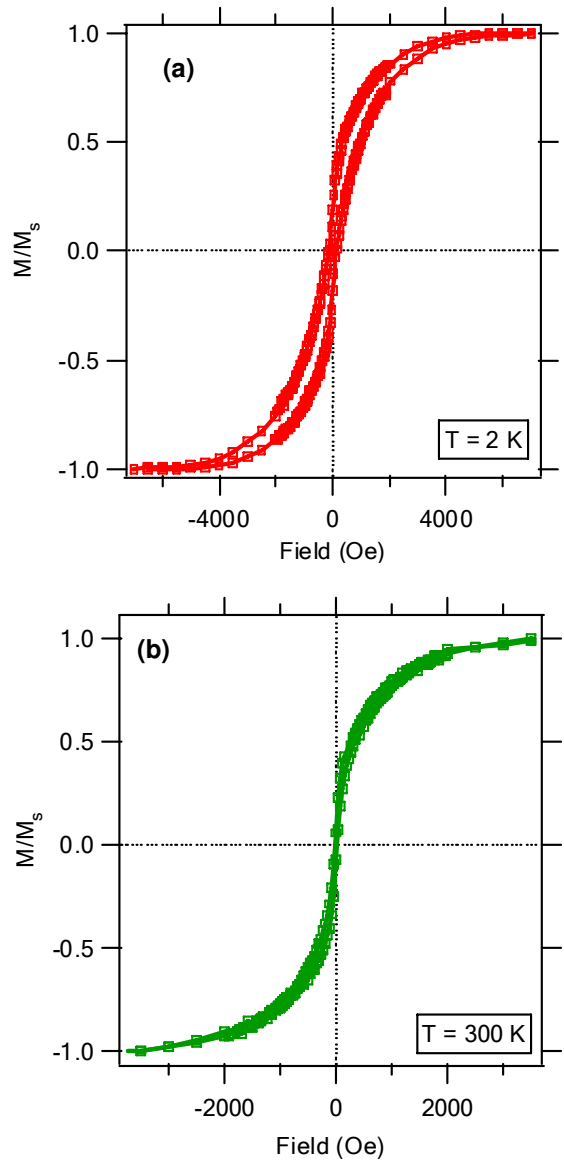} \vspace{-2mm}
\caption{(color online) (a) Magnetization measurements data for
thin wall small rings at 2 K. (b) Magnetic measurement at 300 K. }
\vspace{-4mm}
\end{figure}

In this approach, a reasonable model for magnetization
distribution on ring's circumference is assumed in zero field
value and total energy is calculated for different possible
magnetic states, starting from the first principle. For any shape
of body, those magnetic configurations which are energetically
favored are dominant in determining the magnetic
behavior.\cite{Rothman} In general, the magnetic state of a ring
will be determined by the competition between exchange energy,
magnetostatic energy, Zeeman energy and magneto-crystalline
energy. The exchange energy contribution favors the parallel
alignment of the local magnetization $\textbf{m}$ over the entire
body while magnetostatic energy favors configurations where the
magnetization follows a closed path inside the body, so that no
net magnetic moment is produced.

Before considering the magnetization distributions on the ring's
circumference in detail, we first calculate the energies of the
uniformly magnetized SD, flux closure vortex and onion states near
zero field (Figure 1). For this purpose, first we consider
ultra-small ring. Since the thickness (5 nm) and width (4 nm) of
the ultra-small ring are comparable to the exchange length of Co
material ($l$$_{ex}$ = 3.8 nm),\cite{Coey} magneto-crystalline
anisotropy can be ignored for simplicity. For an ultra-small ring,
the energetic analysis described below shows that only single
domain and vortex states are of significance at fields near zero;
the energy of the onion state lies far above. An important
underlying assumption in the calculation of energy is that since
the thickness and width of the ring are comparable to the exchange
length for cobalt, there is no variation of magnetization along
the symmetry axis (easy axis), of the ring i.e. magnetization
pattern is purely two-dimensional in nature. Recently other
authors\cite{Kravchuk}$^{,}$ \cite{Landeros} have found that the
variation of magnetization along the symmetry axis in the
relatively larger size magnetic rings leads to triple point
behavior, separating SD, vortex and onion magnetic states as a
function of material's properties, in phase diagram. We have
discussed these results later in the discussion section.

In dimensionless form, the free energy for a magnetic system of
volume $V$ is given by,\cite{Bertotti}
\begin{eqnarray*}
{E}&=&{E}_{ex}+{E}_{an}+{E}_{m}+{E}_{zeeman}\\
   &=&\frac{1}{V}{\int}(\frac{{l}_{ex}^{2}}{2}({\nabla}{m}^{2})+{\kappa}{f}({m})-\frac{1}{2}({h}_{m}{\bullet}{m})-{h}_{a}{\bullet}{m}){d}^{3}{r}
\end{eqnarray*}
Here $h$$_{m}$ = $H$$_{M}$/$M$$_{s}$ and $h$$_{a}$ =
$H$$_{a}$/$M$$_{s}$ are dimensionless magnetic fields. $H$$_{a}$
is the applied magnetic field, and $H$$_{M}$ is the magnetic field
produced by the magnetization of the sample, which has saturation
magnetization $M$$_{s}$. $\kappa$ =
2$K$$_{1}$/$\mu$$_{0}$$M$$_{s}$$^{2}$ = $H$$_{an}$/$M$$_{s}$,
where $K$$_{1}$ is the first crystalline anisotropy constant
(which will be taken to be zero for simplicity). The exchange
length $l$$_{ex}$ = $\surd$(2$A$/$\mu$$_{0}$$M$$_{s}$$^{2}$),
where $A$ is the exchange stiffness constant. For the uniformly
magnetized SD state, the exchange energy is zero and the total
energy is the sum of the magnetostatic and Zeeman contributions.
\begin{eqnarray}
{E}_{SD}&=&{E}_{m}+{E}_{zeeman}\\
        &=& 0.2-{h}_{a}
\end{eqnarray}

In the above equation, the dimensionless magnetostatic energy
$E$$_{m}$= 0.20 was obtained by numerical integration. Numerical
integration for magnetostatic energy is carried out by assuming
the ring's inner and outer surfaces as two oppositely magnetic
charged ribbons separated from each other by the width of rings.
Thickness of that ribbon is 5 nm (ring's thickness). We write the
magnetostatic potential at an arbitrary point, in the region
between ribbons, and then numerically integrate it over the volume
of the ring.

For the vortex state, the magnetostatic, domain wall and Zeeman
energy contributions are all zero, so the total energy is solely
exchange energy.
\begin{eqnarray}
{E}_{vortex}&=&{E}_{ex}\\
            &=&\frac{{l}_{ex}^{2}}{{R}_{2}^{2}-{R}_{1}^{2}}[{\ln}\frac{{R}_{2}}{{R}_{1}}]
\end{eqnarray}
To calculate the energy of the onion configuration, we separate it
into two parts: the ring arms and the domain walls (DW). The
contributing energy terms are exchange, Zeeman and magnetostatic.
Exchange energy is proportional to $\nabla$$\textbf{m}$$^{2}$, so
the vortex and the ring arms of the onion state have the same
exchange energy. The energy of onion magnetic state is
\begin{eqnarray*}
{E}_{onion}&=&{E}_{ex}+{E}_{zeeman}+{E}_{m(ringarms)}+{E}_{DW}\\
           &=&\frac{{l}_{ex}^{2}}{{R}_{2}^{2}-{R}_{1}^{2}}{\ln}(\frac{{R}_{2}}{{R}_{1}})-\frac{2}{\pi}[{h}_{a}]+{E}_{m(ringarms)}+{E}_{DW}
\end{eqnarray*}
where $R$$_{1}$ is the inner radius, $R$$_{2}$ is the outer
radius. Both $E$$_{m(ringarms)}$ and $E$$_{DW}$ are positive: each
increases the total energy of the onion state.

\begin{figure}
\centering
\includegraphics[width=8.0cm]{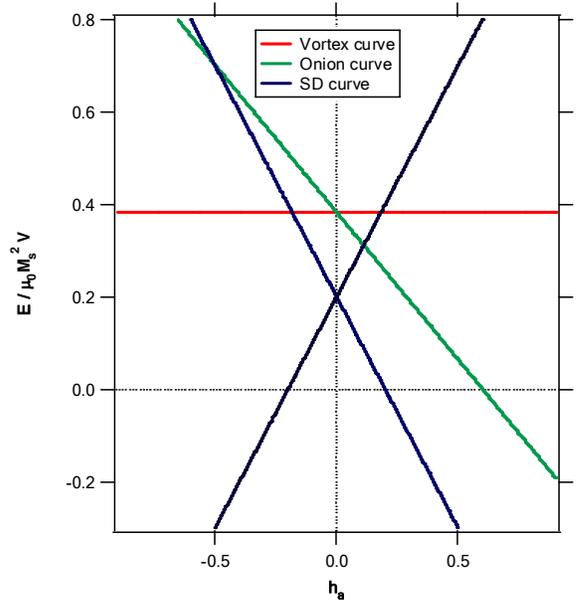} \vspace{-2mm}
\caption{(color online) Magnetic energies of ultra-small rings.
Energies of SD, vortex and onion states as functions of
dimensionless applied field $h$$_{a}$. (The onion curve is a lower
limit - see text.)
} \vspace{-4mm}
\end{figure}

The energies of the SD and vortex states, together with a lower
limit of the onion state energy, are plotted as functions of the
dimensionless applied field $h$$_{a}$ in Figure 5. The lower limit
is obtained by neglecting the domain wall and magnetostatic terms.
This figure suggests that the energy landscape has three local
minima and that these three magnetic states have locally minimal
energies. According to this energy plot, it is inferred that SD
state is always lowest in energy. Vortex and onion magnetic state
does not excite. Hence the corresponding magnetic hysteresis curve
would be a one step transition, from one SD state to another SD
state of opposite polarity. This is not exactly the experimental
data for ultra-small magnetic rings (Fig. 3).

As mentioned previously, rings are made of polycrystalline Co
material. In the case of polycrystalline material, random
anisotropies of grains play a crucial role in determining the
intergranular interaction. If the grains are small then
intergranular interactions becomes more dominating than
interatomic exchange interaction and thus lower the exchange
length values. As a result, the energy of vortex magnetic state
would be smaller. In that case, a transition from saturating SD
state to vortex state, near zero field value, is quite likely.

Depending upon the polycrystalline exchange length of Co, magnetic
transition in ultra-small ring can occur from SD state to vortex
state via the distortion of pure SD state, or from the pure vortex
state to pure SD state via the distortion of vortex state. An
important question arises then: are these intermediate (distorted)
magnetic states stable i.e. lower in energy compared to pure SD
and vortex states respectively. It is found that small distortions
of the SD state increases its energy. In the following section,
detail calculation and results are explained for distorted SD and
vortex state.

\section{DISTORTED SINGLE DOMAIN AND VORTEX STATES}
For an ultra-small ring, at very high field the magnetic state is
saturated single domain. When the field is reduced from the
saturation value, SD state is supposed to distort and eventually
attains a vortex configuration near zero value of field, depending
on the polycrystalline exchange length of Co material. A schematic
description of this phenomenon is shown in Figure 6. In this
figure, we call the magnetic state at intermediate field value as
the "distorted single domain state". If the distorted SD state is
unfolded into one dimension, then the envelop over magnetic spin
shows an oscillatory character, as shown below in Figure 7a.

\begin{figure}
\centering
\includegraphics[width=8.0cm]{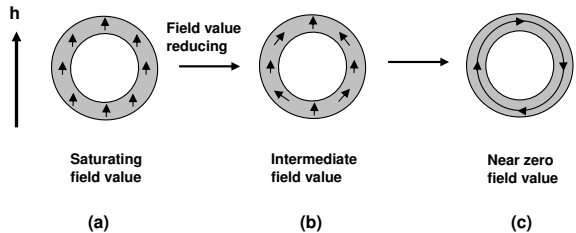} \vspace{-2mm}
\caption{(color online) (a) At very high field value, the magnetic
state of ring is single-domain, (b) when the field is reduced then
possible magnetic state may be a distorted SD state and (c) near 0
field value, the magnetic state of ring becomes vortex state. }
\vspace{-4mm}
\end{figure}

\begin{figure}
\centering
\includegraphics[width=8.0cm]{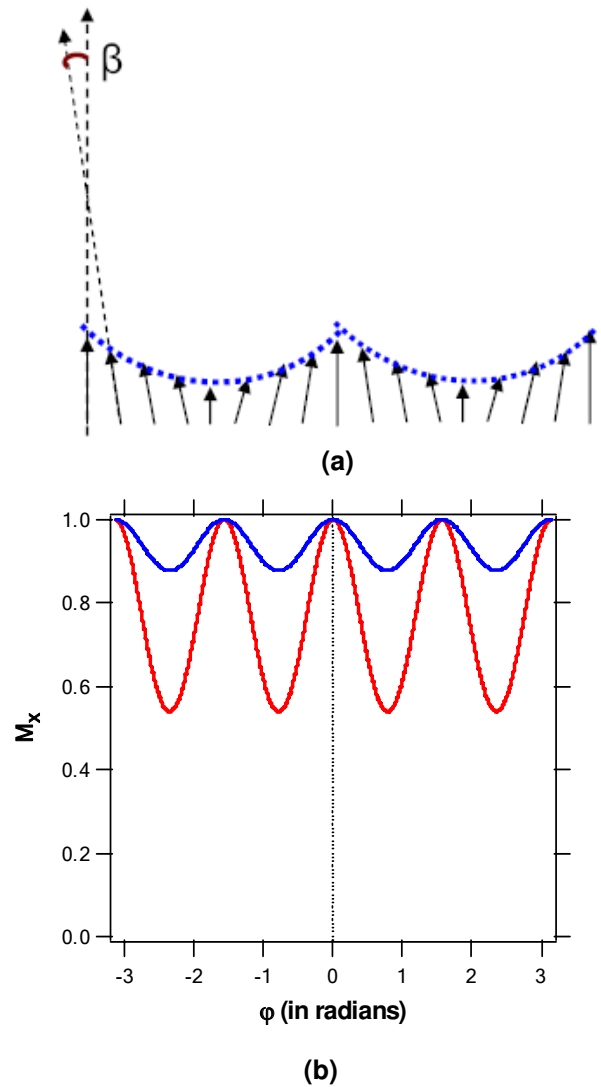} \vspace{-2mm}
\caption{(color online) (a) One dimensional profile of magnetic
spins of distorted SD state. Magnetic spins are represented by the
azimuthal angle $\phi$ and angle between one arbitrary spin and
the applied field axis is $\beta$. In this model angle ? is
supposed to be slowly varying through the circumference of ring.
(b) In this picture variation of $m$$_{x}$ is plotted as a
function of $\phi$ for two different values of distortion
parameters $\varepsilon$ = 0.2 (Blue curve) and 0.5 (Red curve) }
\vspace{-4mm}
\end{figure}

In Figure 7a, $\beta$ is the angle between an arbitrary spin and
the applied field axis. An arbitrary spin is represented by the
angle $\phi$ on the circumference of ring. So $\phi$ varies
between 0 and 2$\pi$. The very important underlying assumption in
this model is the slow variation of angle $\beta$ over the
circumference of ring in the distorted single domain magnetic
state. Now we seek a function $\beta$ = $f$($\phi$), which will
give us magnetization distribution as shown in the profile in
Figure 7a. By keeping in mind that the profile of magnetic spins
shows oscillatory behavior, we make an assumption for the
variation of angle $\beta$ as a function of $\phi$ as:
\begin{eqnarray}
{\beta}&=&{\varepsilon}{sin}{2}{\phi}
\end{eqnarray}
\begin{eqnarray}
{m}_{x}&=&{cos}{\beta}\\
       &=&{cos}(-{\varepsilon}{sin}{2}{\phi})
\end{eqnarray}
where $\varepsilon$ is the distortion coefficient and $X$ is the
direction of externally applied magnetic field. Figure 7b shows
the variations of $m$$_{x}$ as a function of $\phi$ for two
different distortion parameters of $\varepsilon$ = 0.2 and 0.5. It
is a reasonably good approximation (if not exact) for
magnetization distribution. For different values of distortion
coefficients, the distortion of magnetic spins on the
circumference of a ring is shown in Figure 8.

\begin{figure}
\centering
\includegraphics[width=8.0cm]{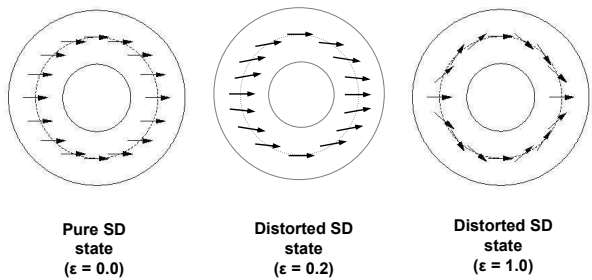} \vspace{-2mm}
\caption{(color online) (a) In this picture we see that for the
value of distortion coefficient $\varepsilon$ = 0, all the spins
are aligned in along the magnetic field direction and thus
creating a SD magnetic state. (b) For the value of distortion
coefficient $\varepsilon$ = 0.2, not all the spins are aligned
along the magnetic field direction and therefore creating a
distorted SD state. (c) Highly distorted SD state, $\varepsilon$ =
1.0.
} \vspace{-4mm}
\end{figure}

In single domain state, $\beta$($\phi$) = 0, so $\varepsilon$ = 0,
which is same as in Figure 8. On the other hand if $\varepsilon$ =
1 then not all the spins are aligned along the magnetic field.
Spins which are not aligned along the magnetic field appear to be
distorted from their original direction while other spins are
still aligned along the magnetic field direction. We call this
state as distorted SD magnetic state. At zero applied field, the
total dimensionless energy, g($\varepsilon$), of distorted SD
state is the sum of two terms:

g($\varepsilon$) = g$_{ex}$($\varepsilon$)+g$_{ms}$($\varepsilon$)

As mentioned in the previous section, we reasonably ignore
crystalline anisotropy term in total energy calculation.

Starting from the first principle (as described in previous
section), exchange energy and magnetostatic energy are calculated
as a function of distortion coefficient ($\varepsilon$) for
ultra-small ring geometry. All the three energy terms, exchange
energy, magnetostatic energy and total energy are plotted in
Figure 11 as a function of distortion coefficient $\varepsilon$.
In this figure, if all the spins are aligned in one direction i.e.
no distortion ($\varepsilon$ = 0) and thus creating a SD state
then exchange energy is minimum (zero) but magnetostatic energy is
maximum. As the distortion is increased so that spins are not
aligned in one direction and therefore it is no longer a pure SD
state then exchange energy also increases for increasing values of
$\varepsilon$ but magnetostatic energy shows different trend.
Magnetostatic energy starts decreasing for increasing values of
distortion coefficient but for higher distortions ($\varepsilon$
$\geq$ 1.2) the magnetostatic energy starts increasing again. The
total energy of distorted SD state increases with increasing value
of $\varepsilon$.

\begin{figure}
\centering
\includegraphics[width=8.0cm]{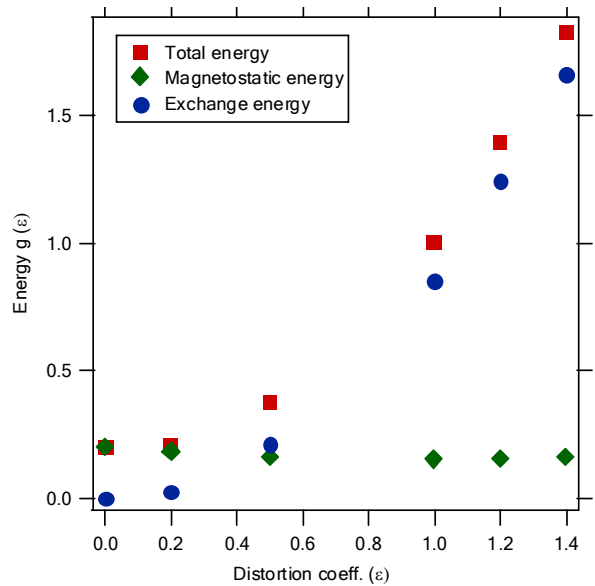} \vspace{-2mm}
\caption{(color online) Energies of pure SD state and distorted SD
state for an ultra-small magnetic ring are plotted in this figure
as a function of distortion coefficient $\varepsilon$. }
\vspace{-4mm}
\end{figure}

Although magnetostatic energy decreases for increasing values of
$\varepsilon$ but it does not affect the overall increase of total
energy. This happens because of larger exchange energy
contributions as the pure SD state becomes more and more
distorted. Thus we conclude that for perturbations of this
particular form, the minimum energy state is still of pure SD
state. It means, for this model function ($\beta$ = $f$($\phi$)),
pure SD state has lower energy than distorted single domain states
in zero field. Since exchange and magnetostatic energies are
independent of field, so even in the presence of field pure SD
state is still of lowest energy as compared to distorted SD
states. For different values of $\varepsilon$ (pure SD and
perturbed SD states), total energy of an ultra-small ring are
plotted in Figure 10 as a function of applied magnetic field. Here
it is important to mention that the distorted SD state with
$\varepsilon$ = 1.0 is not an "onion" magnetic state. The onion
state will have extra energy as a result of formation of domain
walls. Our calculations suggest that distorted states with
$\varepsilon$ greater than zero are higher in energy than the pure
SD state, even though states with different $\varepsilon$ have
different sensitivities to applied field (i.e. Zeeman energy
depends on $\varepsilon$). Perhaps the magnetization does distort,
but to some other shape, that we did not take into account.

\begin{figure}
\centering
\includegraphics[width=8.0cm]{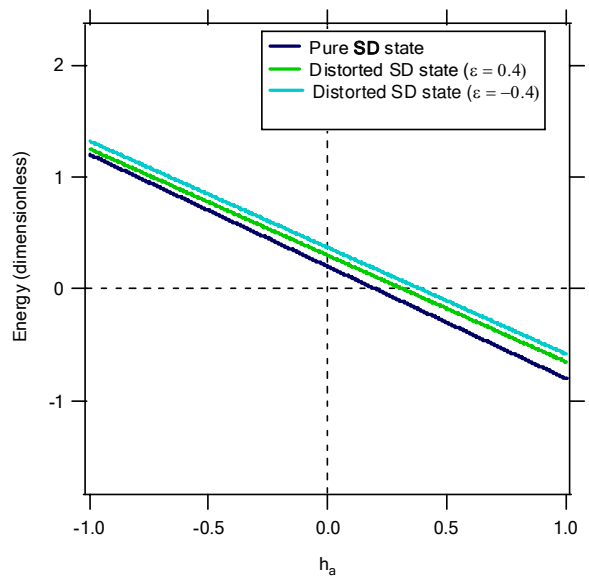} \vspace{-2mm}
\caption{(color online) In this figure, energies of pure SD state
and distorted SD states are plotted as a function of field. }
\vspace{-4mm}
\end{figure}

Another possible magnetic transition involves the magnetic state
of ultra-small ring is already a vortex state near zero value of
field and it makes transition to saturating single domain state
via some intermediate magnetic state. For this purpose, we
consider another realistic model that involves a vortex distortion
constant, $\varepsilon$$_{v}$. In this model, vortex magnetization
state is given by,
\begin{eqnarray*}
{m}_{x}&=&{cos}{\beta}
\end{eqnarray*}
\begin{eqnarray}
{\beta}({\phi})&=&\frac{\pi}{2}+{\phi}-{\varepsilon}_{v}{cos}{\phi}
\end{eqnarray}
once again $\beta$($\phi$) is the angle between the direction of
the magnetization and the field direction (applied along the
x-axis). Different vortex states are shown in Figure 11 for the
vortex distortion constants, $\varepsilon$$_{v}$ , of 0 and 0.5.
For the vortex distortion constant value of $\varepsilon$$_{v}$ =
0 means no distortion and in that case, $\beta$($\phi$) = $\pi$/2
+ $\phi$. We see that for $\varepsilon$$_{v}$ = 0, magnetization
configuration is indeed of pure vortex state. Again, the total
energy of the ultra-small ring in zero field consists of exchange
and magnetostatic energy terms (ignoring the magneto-crystalline
anisotropy energy term in the total energy calculation). The first
principle calculation of exchange energy for distorted vortex
states as a function of $\varepsilon$$_{v}$ is given by the
following equation:
\begin{eqnarray}
{g}_{ex}({\varepsilon}_{v})&=&\frac{{(1/2)}{l}_{ex}^{2}}{{\pi}({R}_{2}^{2}-{R}_{1}^{2})}{\pi}{(2+{\varepsilon}_{v}^{2})}{\ln}\frac{{R}_{2}}{{R}_{1}}
\end{eqnarray}

\begin{figure}
\centering
\includegraphics[width=8.0cm]{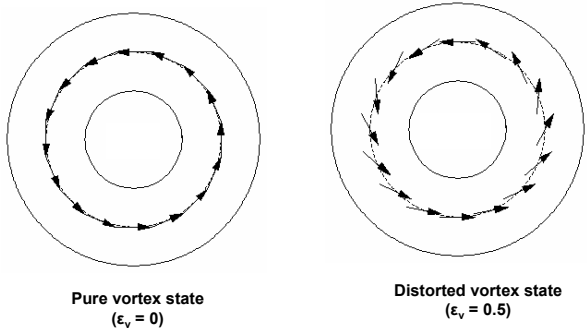} \vspace{-2mm}
\caption{(color online) (a) In this picture we see that for the
value of distortion coefficient $\varepsilon$$_{v}$ = 0,
magnetization spins form a circular vortex magnetic state. (b) For
the value of distortion coefficient $\varepsilon$$_{v}$, all
magnetization spins are not along the tangential direction on a
circle's periphery inside the ring and therefore do not form a
pure vortex state. We call this state a distorted vortex state. }
\vspace{-4mm}
\end{figure}

Magnetostatic energies for distorted vortex states are calculated
by numerical integrations (as explained previously). The energies
for distorted vortex states as a function of vortex distortion
constant $\varepsilon$$_{v}$ are plotted in Fig. 12, in zero
applied field. In this figure, the pure vortex state is indeed the
lowest energy state in zero field because of the dominance of
exchange energy term in the total energy. However as we will see
in the following paragraph, this is not the case in the presence
of field.

\begin{figure}
\centering
\includegraphics[width=8.0cm]{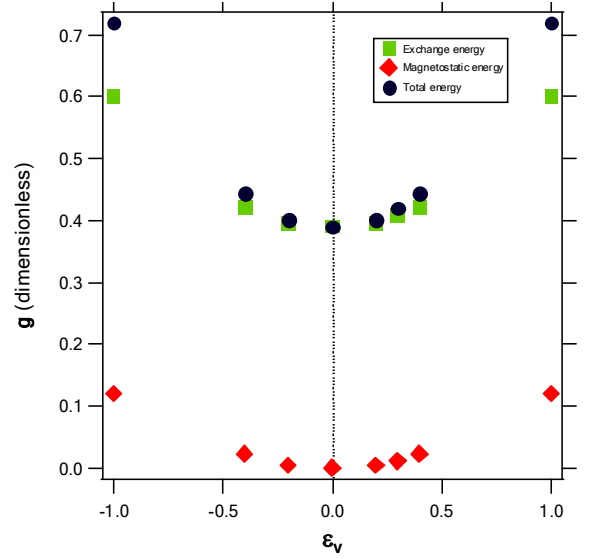} \vspace{-2mm}
\caption{(color online) Energies of pure and distorted vortex
states, in zero field, are plotted in this figure as a function of
distortion coefficient $\varepsilon$$_{v}$.
} \vspace{-4mm}
\end{figure}

In the presence of field, the total energy consists of three
terms:

g($\varepsilon$$_{v}$) = g$_{ex}$($\varepsilon$$_{v}$) +
g$_{ms}$($\varepsilon$$_{v}$) + g$_{zeeman}$($\varepsilon$$_{v}$)

The calculated Zeeman energies for distorted vortex states is
given by:
\begin{eqnarray*}
{g}_{zeeman}({\varepsilon}_{v})&=&-\frac{{h}_{a}}{{2}/{\pi}}{2}{\pi}{Bessel}{J}[1,{abs}({\varepsilon}_{v})]{sign}[{\varepsilon}_{v}]
\end{eqnarray*}
The Zeeman energy is proportional to the applied field $h$$_{a}$ .
So at first instance it appears that pure vortex state will always
remain the magnetic state of lowest energy but at the same time we
see from the Zeeman energy expression that the slope of the Zeeman
energy curve depends on the distortion parameter
$\varepsilon$$_{v}$. Total energy, including the Zeeman energy,
for pure as well as distorted vortex states (distortion parameters
$\varepsilon$$_{v}$ = 0.2, 0.3) are plotted in Figure 13 as a
function of applied field. Some very interesting behaviors are
observed in this figure.

\begin{figure}
\centering
\includegraphics[width=8.0cm]{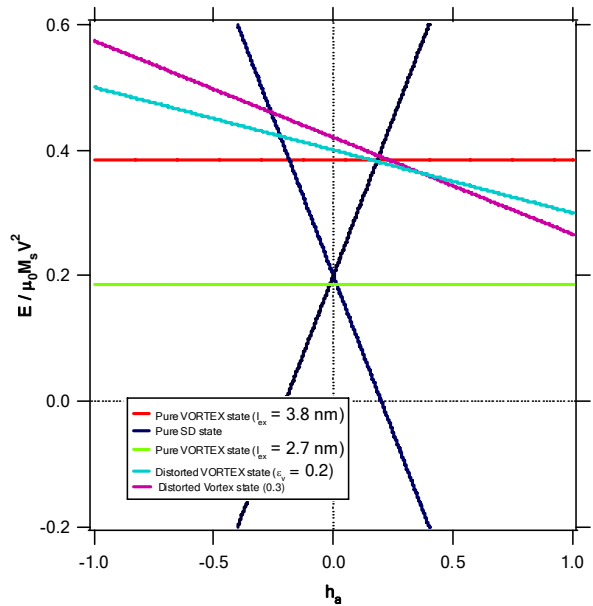} \vspace{-2mm}
\caption{(color online) This plot shows total magnetic energies of
pure and perturbed vortex magnetic configurations (red, light blue
and green lines), together with the energy of the pure SD state
(dark blue line) as a function of applied field. Pure SD state is
always the lowest energy state.
} \vspace{-4mm}
\end{figure}

Near zero field values, pure vortex state is still the lowest
energy state. At a field of 0.2 or so the state with distortion
$\varepsilon$$_{v}$ = 0.2 crosses over to become lower than the
state with distortion $\varepsilon$$_{v}$ = 0.0. At a somewhat
higher field the green state with distortion parameter
$\varepsilon$$_{v}$ = 0.3 becomes the one with lowest energy. This
is not exactly the case for distorted SD states (Figure 10), as
discussed earlier. In the case of single domain states, pure SD
state remains lowest in energy as the applied field is changed.
Therefore distorted single domain states can be ignored in
considering the magnetic transition for ultra-small rings. Thus
there are three important states to consider: pure SD state, pure
vortex state and distorted vortex state. In Figure 13 we have
plotted all these energies as a function of applied field. It is
observed that pure SD state is always the lowest energy state. In
this figure, an additional vortex state, arising for a lower
exchange length value ($l$$_{ex}$ = 2.7 nm) of polycrystalline Co
material, is also plotted. This vortex curve intercepts the pure
SD curves at dimensionless $h$$_{a}$ = $\pm$ 0.01. The
implications are explained in the following section.

\section{DISCUSSION}
Based on the above analysis and crystalline exchange length value
($l$$_{ex}$ = 3.8 nm), the magnetization hysteresis curve for
ultrasmall ring would be one step transition at zero field which
is not quite the experimental data (Figure 3). The experimental
hysteresis curve shows zero magnetization at $h$$_{a}$ = $\pm$0.01
(equivalent to $\pm$ 90 Oe of applied magnetic field). Now, as
mentioned previously, the rings are poly-crystalline, resulting in
an effective exchange length that is smaller than the 3.8 nm bulk
value used in the calculations. For a given geometry, the energy
of a vortex at zero field is proportional to the square of the
exchange length, while the energy of an SD state is independent of
the exchange length. If we consider an exchange length of 2.7 nm
for the polycrystalline Co material then the vortex state has the
lowest energy at zero field and is degenerate with an SD state at
fields of about $h$$_{a}$ = $\pm$0.01 (Figure 13). Theoretically,
the corresponding hysteresis curve would be two steps separated by
a short plateau of width about 0.02, as shown in Figure 13. This
provides a qualitative explanation of the experimental
magnetization data of ultra-small magnetic rings at $T$ = 2 K.

From this analysis we can conclude that near zero field,
ultra-small rings will be in either a single domain state or a
vortex state, depending upon the polycrystalline exchange length
value. The onion state is not expected to play a role.

\begin{figure}
\centering
\includegraphics[width=8.0cm]{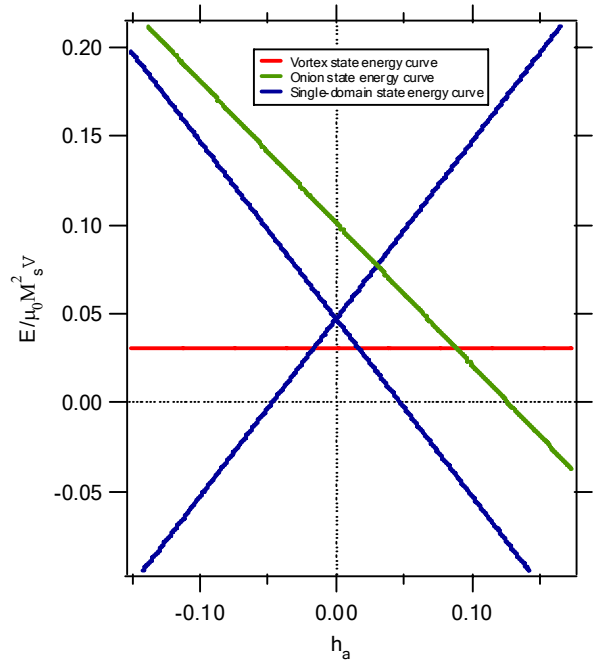} \vspace{-2mm}
\caption{(color online) Energy plot for thin wall small ring of
diameter 150 nm, width 5 nm and thickness 5nm is shown in this
figure. In the case of small ring of thin walls, the magnetic
transition occurs via the formation of vortex state, even for the
exchange length value of $l$$_{ex}$ = 3.8 nm.
} \vspace{-4mm}
\end{figure}

Now we check the validity of above analysis for ultra-small
magnetic rings in relatively longer geometrical regime of small
ring of thin wall (outer diameter 150 nm). As mentioned in the
fabrication section, small ring's width and thicknesses are
comparable to that of ultra-small rings and therefore to the
characteristic length of Co. However, the outer diameter of small
rings are an order of magnitude larger than the diameter of
ultra-small rings. The vortex energy and magnetostatic energy
depends on the outer diameter of the ring geometry. Based on the
first principle calculation, the three possible energy terms SD,
onion and vortex magnetic state are plotted as a function of
magnetic field in Figure 14. As inferred from Fig. 14, in this
case also the onion state does not arise and the magnetic
transition occurs between two SD states via the formation of
vortex magnetic state. Interestingly, in the case of thin wall
small rings vortex magnetic state arises even for the exchange
length value of $l$$_{ex}$ = 3.8 nm. The reason behind this is the
dominance of magnetostatic energy term in the overall energy
counting. Therefore as the lateral size of a ring increases,
magnetostatic energy starts dominating over the exchange energy
term. The experimental (in-plane) magnetization data (at 2 K) for
small rings, Figure 4, does not seem to be in good agreement with
Fig. 14. Although the magnetic transition starts occurring around
$\sim$ 1000 Oe of field value but the curvature of magnetic
transition is very large, in terms of field, indicating that the
magnetic transition process in small rings may not be as simple.
As the overall ring size increases, the magnetostatic energy of
the distorted SD states start dominating in the total energy
counting and therefore enhances the possibility of interaction
between individual ring elements. It leads to magnetocrystalline
anisotropy in the ring arrays that we have not taken into account
in the calculation. Some authors have calculated anisotropy energy
for relatively thicker magnetic rings.\cite{Zhu3} More general
first principle calculation would be necessary to understand the
magnetization reversal in thin wall small rings.

\begin{figure}
\centering
\includegraphics[width=8.0cm]{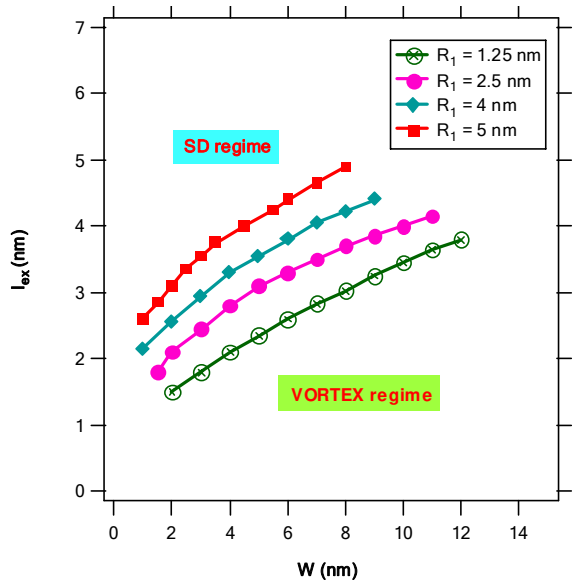} \vspace{-2mm}
\caption{(color online) This figure represents the phase-diagram
for ultra-small magnetic rings with in-plane (2-D) magnetization
pattern. It also shows that whether SD or vortex state is lowest
at zero field for a given size ring.
} \vspace{-4mm}
\end{figure}

Magnetic measurements at room temperature for both ultra-small and
small rings do not show any magnetic hysteresis. At high
temperature, thermal fluctuations play dominant role, as compared
to the magnetic energy of these rings. This cancels out any
remnant magnetization at zero field. Weakly noticeable
asymmetricity of magnetization loops for positive and negative
fields, in the low temperature magnetic hysteresis curve of
ultra-small ring arrays (Fig. 3a), is possibly due to the weak
exchange bias phenomena. On the surface of the ring, partial
oxidization of Co material into CoO creates an interface of FM
(Co) / AFM (CoO) layer, resulting into very weak exchange bias
phenomena. A similar behavior has recently been reported in the
case of magnetic disk.\cite{Sort}

Now we summarize the above analysis for ultra-small ring geometry.
Rings made of polycrystalline Co material can have smaller
exchange length than the crystalline value. This exchange length,
along with the ring's geometrical parameters, decides which
magnetic state has the lowest energy at zero field. Increasing the
diameter reduces both the magnetostatic and exchange energies but
the exchange energy decreases more rapidly and therefore
magnetostatic energy starts dominating as the overall diameter of
the magnetic ring increases. For given inner and outer radii
difference, there is a value of the exchange length at which the
vortex and SD states are degenerate at zero field. We have also
drawn a phase-diagram in Figure 15 that shows the value of
exchange length, $l$$_{ex}$ as a function of ring width $w$
($R$$_{2}$ - $R$$_{1}$). If the ring's exchange length is larger
than the ordinate of that point, the SD states will be lowest at
zero field, otherwise a vortex state will be the lowest. This
phase diagram may not be valid for larger size nanoscopic rings.
As there is a strong underlying assumption in the above analysis
is that the magnetization pattern in the case of ultra-small
magnetic ring is purely 2-D. Magnetization vector does not
cant-out of the ring's plane. For an ultra-small ring with width
and thicknesses comparable to the characteristic length (exchange
length) of the parent magnetic material, this assumption is quite
reasonable. As a consequence, we do not observe the "triple point"
-- defined as the point where the energies for onion, vortex and
uniform out-of-plane magnetization are same -- as recently
reported by other authors based on theoretical calculations for
the ring geometry.\cite{Kravchuk}$^{,}$ \cite{Landeros}.

\section*{Acknowledgements}
This project was supported by NSF Grants DMR-0531171, DMR-0306951
and MRSEC.

\bibliography{Ring2}

\begin{thebibliography}{26}
\expandafter\ifx\csname natexlab\endcsname\relax\def\natexlab#1{#1}\fi
\expandafter\ifx\csname bibnamefont\endcsname\relax
  \def\bibnamefont#1{#1}\fi
\expandafter\ifx\csname bibfnamefont\endcsname\relax
  \def\bibfnamefont#1{#1}\fi
\expandafter\ifx\csname citenamefont\endcsname\relax
  \def\citenamefont#1{#1}\fi
\expandafter\ifx\csname url\endcsname\relax
  \def\url#1{\texttt{#1}}\fi
\expandafter\ifx\csname urlprefix\endcsname\relax\def\urlprefix{URL }\fi
\providecommand{\bibinfo}[2]{#2}
\providecommand{\eprint}[2][]{\url{#2}}

\bibitem[{\citenamefont{Yoo\emph{, et al.}}(2003)}]{Yoo}
\bibinfo{author}{\bibfnamefont{Y.~G.} \bibnamefont{Yoo\emph{, et al.}}},
  \bibinfo{journal}{Appl. Phys. Lett.} \textbf{\bibinfo{volume}{82}},
  \bibinfo{pages}{2470} (\bibinfo{year}{2003}).

\bibitem[{\citenamefont{Klaui\emph{, et al.}}(2003)}]{Klaui1}
\bibinfo{author}{\bibfnamefont{M.}~\bibnamefont{Klaui\emph{, et al.}}},
  \bibinfo{journal}{J. Phys.: Cond. Matt.} \textbf{\bibinfo{volume}{15}},
  \bibinfo{pages}{R985} (\bibinfo{year}{2003}).

\bibitem[{\citenamefont{Natali\emph{, et al.}}(2002)}]{Natali}
\bibinfo{author}{\bibfnamefont{M.}~\bibnamefont{Natali\emph{, et al.}}},
  \bibinfo{journal}{Phys. Rev. Lett.} \textbf{\bibinfo{volume}{88}},
  \bibinfo{pages}{157203} (\bibinfo{year}{2002}).

\bibitem[{\citenamefont{Zhu\emph{, et al.}}(2000)}]{Zhu1}
\bibinfo{author}{\bibfnamefont{J.~G.} \bibnamefont{Zhu\emph{, et al.}}},
  \bibinfo{journal}{Jour. Appl. Phys.} \textbf{\bibinfo{volume}{87}},
  \bibinfo{pages}{6668} (\bibinfo{year}{2000}).

\bibitem[{\citenamefont{Rothman\emph{, et al.}}(2001)}]{Rothman}
\bibinfo{author}{\bibfnamefont{J.}~\bibnamefont{Rothman\emph{, et al.}}},
  \bibinfo{journal}{Phys Rev Lett} \textbf{\bibinfo{volume}{86}},
  \bibinfo{pages}{1098} (\bibinfo{year}{2001}).

\bibitem[{Sko()}]{Skomsky}
\bibinfo{note}{R. Skomsky, Chapter 10, Spin Elctronics, Springer Publishing
  Group (2000)}.

\bibitem[{\citenamefont{Cowburn\emph{, et al.}}(2000)}]{Cowburn1}
\bibinfo{author}{\bibfnamefont{R.~P.} \bibnamefont{Cowburn\emph{, et al.}}},
  \bibinfo{journal}{J. Phys.D: Appl. Phys.} \textbf{\bibinfo{volume}{33}},
  \bibinfo{pages}{R1} (\bibinfo{year}{2000}).

\bibitem[{\citenamefont{Aharoni\emph{, et al.}}(1992)}]{Aharoni}
\bibinfo{author}{\bibfnamefont{A.}~\bibnamefont{Aharoni\emph{, et al.}}},
  \bibinfo{journal}{Phys. Rev. B} \textbf{\bibinfo{volume}{45}},
  \bibinfo{pages}{1030} (\bibinfo{year}{1992}).

\bibitem[{Coe()}]{Coey}
\bibinfo{note}{J. M. D. Coey, Chapter 12, Spin Elctronics, Springer Publishing
  Group (2000)}.

\bibitem[{Dee()}]{Deepak1}
\bibinfo{note}{Deepak Kumar Singh, Ph. D. Thesis 2006, University of
  Massachusetts Amherst, Available at:
  http://scholarworks.umass.edu/dissertations/AAI3242328/}.

\bibitem[{\citenamefont{Heyderman\emph{, et al.}}(2003)}]{Heyderman}
\bibinfo{author}{\bibfnamefont{L.~J.} \bibnamefont{Heyderman\emph{, et al.}}},
  \bibinfo{journal}{Jour. Appl. Phys.} \textbf{\bibinfo{volume}{93}},
  \bibinfo{pages}{10011} (\bibinfo{year}{2003}).

\bibitem[{\citenamefont{Zhu\emph{, et al.}}(2004)}]{Zhu2}
\bibinfo{author}{\bibfnamefont{F.~Q.} \bibnamefont{Zhu\emph{, et al.}}},
  \bibinfo{journal}{Advanced Materials} \textbf{\bibinfo{volume}{16}},
  \bibinfo{pages}{2155} (\bibinfo{year}{2004}).

\bibitem[{\citenamefont{Hobbs\emph{, et al.}}(2004)}]{Hobbs}
\bibinfo{author}{\bibfnamefont{K.~L.} \bibnamefont{Hobbs\emph{, et al.}}},
  \bibinfo{journal}{NanoLetter} \textbf{\bibinfo{volume}{4}},
  \bibinfo{pages}{167} (\bibinfo{year}{2004}).

\bibitem[{\citenamefont{Kosiorek\emph{, et al.}}(2005)}]{Kosiorek}
\bibinfo{author}{\bibfnamefont{A.}~\bibnamefont{Kosiorek\emph{, et al.}}},
  \bibinfo{journal}{SMALL} \textbf{\bibinfo{volume}{1}}, \bibinfo{pages}{439}
  (\bibinfo{year}{2005}).

\bibitem[{\citenamefont{Singh\emph{ et al.}}(2008)}]{Deepak2}
\bibinfo{author}{\bibfnamefont{D.~K.} \bibnamefont{Singh\emph{ et al.}}},
  \bibinfo{journal}{Nanotechnology} \textbf{\bibinfo{volume}{19}},
  \bibinfo{pages}{245305} (\bibinfo{year}{2008}).

\bibitem[{\citenamefont{Albretch\emph{ et al.}}(2004)}]{Albretch}
\bibinfo{author}{\bibfnamefont{T.}~\bibnamefont{Albretch\emph{ et al.}}},
  \bibinfo{journal}{Advanced Materials} \textbf{\bibinfo{volume}{12}},
  \bibinfo{pages}{787} (\bibinfo{year}{2004}).

\bibitem[{\citenamefont{Bal\emph{ et al.}}(2002)}]{Bal}
\bibinfo{author}{\bibfnamefont{M.}~\bibnamefont{Bal\emph{ et al.}}},
  \bibinfo{journal}{Appl. Phys. Lett.} \textbf{\bibinfo{volume}{81}},
  \bibinfo{pages}{3479} (\bibinfo{year}{2002}).

\bibitem[{\citenamefont{Klaui\emph{ et al.}}(2004)}]{Klaui2}
\bibinfo{author}{\bibfnamefont{M.}~\bibnamefont{Klaui\emph{ et al.}}},
  \bibinfo{journal}{Appl. Phys. Lett.} \textbf{\bibinfo{volume}{85}},
  \bibinfo{pages}{5637} (\bibinfo{year}{2004}).

\bibitem[{\citenamefont{Li\emph{, et al.}}(2001)}]{Li}
\bibinfo{author}{\bibfnamefont{S.~P.} \bibnamefont{Li\emph{, et al.}}},
  \bibinfo{journal}{Phys Rev Lett} \textbf{\bibinfo{volume}{86}},
  \bibinfo{pages}{1102} (\bibinfo{year}{2001}).

\bibitem[{Bro()}]{Brown}
\bibinfo{note}{"Micromagnetics", Interscience Publishers (1963)}.

\bibitem[{Aha()}]{Aharoni2}
\bibinfo{note}{A. Aharoni, Theory of Ferromangetism, Oxford Science Publication
  (2000)}.

\bibitem[{\citenamefont{Kravchuk\emph{, et al.}}(2006)}]{Kravchuk}
\bibinfo{author}{\bibfnamefont{V.~P.} \bibnamefont{Kravchuk\emph{, et al.}}},
  \bibinfo{journal}{J. Mag. Mag. Mat.} \textbf{\bibinfo{volume}{310}},
  \bibinfo{pages}{116} (\bibinfo{year}{2006}).

\bibitem[{\citenamefont{Landeros\emph{, et al.}}(2007)}]{Landeros}
\bibinfo{author}{\bibfnamefont{P.}~\bibnamefont{Landeros\emph{, et al.}}},
  \bibinfo{journal}{Jour. Appl. Phys.} \textbf{\bibinfo{volume}{100}},
  \bibinfo{pages}{044311} (\bibinfo{year}{2007}).

\bibitem[{Ber()}]{Bertotti}
\bibinfo{note}{M. Bertotti, Hysteresis and magnetism, Academic Press (1998)}.

\bibitem[{\citenamefont{Zhu\emph{, et al.}}(2006)}]{Zhu3}
\bibinfo{author}{\bibfnamefont{F.~Q.} \bibnamefont{Zhu\emph{, et al.}}},
  \bibinfo{journal}{Phys. Rev. Lett.} \textbf{\bibinfo{volume}{96}},
  \bibinfo{pages}{027205} (\bibinfo{year}{2006}).

\bibitem[{\citenamefont{Sort\emph{, et al.}}(2005)}]{Sort}
\bibinfo{author}{\bibfnamefont{J.}~\bibnamefont{Sort\emph{, et al.}}},
  \bibinfo{journal}{Phys. Rev. Lett.} \textbf{\bibinfo{volume}{97}},
  \bibinfo{pages}{067201} (\bibinfo{year}{2005}).

\end{thebibliography}
*email: tuominen@physics.umass.edu
\end{document}